\documentclass[12pt]{article}
\usepackage[onehalfspacing]{setspace}
\usepackage{graphicx}
\usepackage[english]{babel}
\usepackage{amssymb,amsmath,amsthm,amsfonts}
\usepackage{color}
\usepackage{natbib,hyperref}
\usepackage{caption,subfig}

\newcommand{\data}{\pmb{x}}
\newcommand{\dataObs}{\pmb{x^0}}
\newcommand{\datasim}{\pmb{x^{\mbox{sim}}}}
\newcommand{\distance}{d}
\newcommand{\Model}{\mathcal{M}}
\DeclareMathOperator*{\argmin}{arg\,min}
\newcommand{\parameter}{\pmb{\theta}}
\newcommand{\prior}{\pi}
\newcommand{\estparameter}{\hat{\pmb{\theta}}}
\newcommand{\sac}{\mathcal{S}_{agg-clust}}
\newcommand{\nac}{\mathbb{N}_{agg-clust}}
\newcommand{\np}{\mathbb{N}_{platelet}}
\newcommand{\nacp}{\mathbb{N}_{act-platelet}}
\newcommand{\Pg}{p_{Ag}}
\newcommand{\Pad}{p_{Ad}}
\newcommand{\Pt}{p_{T}}
\newcommand{\Pf}{p_{F}}
\newcommand{\Ra}{a_{T}}

\newcommand{\lossfunc}{\mathcal{L}}
\newcommand{\mean}{\mu}
\newcommand{\var}{\sigma}
\newcommand{\autcor}{ac}
\newcommand{\cor}{c}
\newcommand{\crosscor}{cc}
\newcommand{\bc}{\rho}

\def\grad{\mathrm{grad}}
\def\ds{\Delta S}
\def\dt{\Delta t}

\newcommand{\blind}{0}
\begin{document}


\if0\blind
{

\title{\bf Parameter estimation of platelets deposition: Approximate Bayesian computation with high performance computing
} 

\author{Ritabrata Dutta\,$^{1}$, Bastien Chopard\,$^{2}$\thanks{Corresponding author:Bastien.Chopard@unige.ch}, Jonas L\"att\,$^{2}$,\\
Frank Dubois\,$^{5}$, Karim Zouaoui Boudjeltia\,$^{3,4}$ and Antonietta Mira\,$^{1,6}$\\ 
{\em \small $^1$Institute of Computational Science, Universit\`a della Svizzera italiana, Switzerland}\\ 
{\em \small$^2$Computer Science Department, University of Geneva, Switzerland}\\  
{\em \small$^3$Laboratory of Experimental Medicine (ULB 222 Unit), Universit\'e Libre de Bruxelles, Belgium}\\
{\em \small$^4$ CHU de Charleroi, Belgium}\\
{\em \small$^5$ Microgravity Research Centre, Universit\'e libre de Bruxelles (ULB), Belgium} \\ 
{\em \small$^6$Department of Science and High Technology, Universit\`a degli Studi dell'Insubria, Italy}\\
  } \maketitle } \fi

\if1\blind
{
  \bigskip
  \bigskip
  \bigskip
  \begin{center}
    {\LARGE\bf Title}
\end{center}
  \medskip
} \fi

\bigskip
Cardio/cerebrovascular diseases (CVD) have become one of the major health issue in our societies. Recent studies show the existing clinical tests to detect CVD are ineffectual as they do not consider different stages of platelet activation or the molecular dynamics involved in platelet interactions. Further they are also incapable to consider inter-individual variability. 
A physical description of platelets deposition was introduced recently in \cite{Chopard_2017}, by integrating fundamental understandings of how platelets interact in a numerical model, parameterized by five parameters. These parameters specify the deposition process and are relevant for a biomedical understanding of the phenomena. 
One of the main intuition is that these parameters are precisely the information needed for a pathological test identifying CVD captured and that they capture the inter-individual variability. 

Following this intuition, here we devise a Bayesian inferential scheme for estimation of these parameters, using experimental observations, at different time intervals, on the average size of the aggregation clusters, their number per $mm^2$, the number of platelets and the ones activated per $\mu\ell$ still in suspension. 
As the likelihood function of the numerical model is intractable due to the complex stochastic nature of the model, we use a likelihood-free inference scheme approximate Bayesian computation (ABC) to calibrate the parameters in a data-driven manner. As ABC requires the generation of many pseudo-data by expensive simulation runs, we use a high performance computing (HPC) framework for ABC to make the inference possible for this model.

We consider a collective dataset of 7 volunteers and use this inference scheme to get an approximate posterior distribution and the Bayes estimate of these five parameters. The mean posterior prediction of platelet deposition pattern matches the experimental dataset closely with a tight posterior prediction error margin, justifying our main intuition and providing a methodology to infer these parameters given patient data. 
The present approach can be used to build a new generation of personalized platelet functionality tests for CVD detection, using numerical modeling of platelet deposition, Bayesian uncertainty quantification and High performance computing. 

\vspace{1ex}\noindent


\section{Introduction}

Blood platelets play a major role in the complex process of blood
coagulation, involving adhesion, aggregation and spreading
on the vascular wall to stop a hemorrhage while avoiding the vessel
occlusion. Platelets also play a key role in the occurrence of
cardio/cerebro-vascular accidents that constitute a major health issue in our societies. 
In 2015, Cardiovascular diseases (CVD), including disorders of the heart and blood vessels, were the first cause of mortality worldwide, causing 31\% of deaths~\citep{WHO}.
Antiplatelet therapy generally reduces complications in patients undergoing arterial intervention~\citep{Steinhubl:2002,Metha:2001}. However, the individual response to dual antiplatelet therapy is not uniform and consistent studies reported that even under platelets therapy there were recurrences of atherothrombotic events~\citep{Matetzky:2004,Grubel:2005,Geisler:2006,Hochholzer:2006,Marcucci:2009,Price:2009,Sibbing:2009}. 
In most cases, a standard posology is prescribed to patients, which does not take into account the inter-individual variability linked to the absorption or the effectiveness of these molecules. This was supported by a recent study~\citep{koltai:2017}, reporting the high patient-dependency of the response of the antithrombotic drugs. We should also note that the evaluation of the response to a treatment by the existing tests is test-dependent. 

Nowadays, platelet function testing is performed either as an attempt
to monitor the efficacy of anti-platelet drugs or to determine the
cause of abnormal bleeding or pro-thrombotic status. The most common
method consists of using an optical aggregometer that measures the
transmittance of light passing through plasma rich in platelets (PRP)
or whole blood~\citep{Harrison:2009,Born:63}, to evaluate how platelets
tend to aggregate. Other aggregometers determine the amount of
aggregated platelets by electric impedance~\citep{Velik-Salchner:2008}
or luminescence. In specific contexts, flow
cytometry~\citep{Michelson:2002} is also used to assess platelet
reactivity (VASP test~\citep{Bonello:2009}).
Determination of platelet functions using these different existing techniques in patients undergoing coronary stent implantation have been evaluated in~\cite{breet:2010}, which shows the correlation between the clinical biological measures and the occurrence of a cardiovascular event was null for half of the techniques and rather modest for others.
This may be due to the fact that no current test allows the analysis of the different stages of platelet activation or the prediction of the in vivo behavior of those platelets~\citep{picker:11,koltai:2017}. It is well known that the phenomenon of platelet margination (the process of bringing platelets to the vascular wall) is dependent on the number and shape of red blood cells and their flow \citep{piagnerelli:2007}, creating different pathologies for different diseases (e.g., diabetes, End Renal Kidney Disease, hypertension, sepsis).
Further, platelet margination is also known to be influenced by the aspect ratio of surrogate platelet particles \citep{reasor:2013}. 
 Although there is a lot of data reported by recent research works~\citep{maxwell:2007} on the molecules involved in platelet interactions, these studies indicate that there is a lack of knowledge on some fundamental mechanisms that should be revealed by new experiments.

\emph{Hence, the challenge is to find parameters connecting the
  dynamic processes of adhesion and aggregation of platelets to the
  data collected from the individual patients.}  Recently, by
combining digital holography microscopy (DHM) and mathematical
modeling, \cite{Chopard_2015,Chopard_2017} provided a physical
description of the adhesion and aggregation of platelets in the
Impact-R device. A numerical model is developed that quantitatively
describes how platelets in a shear flow adhere and aggregate on a
deposition surface.  This is the first innovation in understanding the
molecular dynamics involved in platelet interactions.  Five parameters
specify the deposition process and are relevant for a biomedical
understanding of the phenomena.  One of the main intuition is that the
values of these parameters (e.g., adhesion and aggregation rates) are
precisely the information needed to assess various possible
pathological situations and quantify their severity regarding
CVD. Further, it was shown in~\cite{Chopard_2017} that, by hand-tuning
the parameters of the mathematical model, the deposition patterns
observed for a set of healthy volunteers in the Impact-R can be
reproduced.

\emph{Assuming that these parameters can determine the severity of CVD, how do we estimate the adhesion and aggregation rates of given patients by a clinical test?}
The determination of these 
adhesion and aggregation rates by hand-tuning is clearly not a solution as we need to search the high-dimensional parameter space of the mathematical model, which becomes extremely expensive and time consuming. We further notice, this has to be repeated for each patient
and thus requires a powerful numerical approach.
In this work, we resolve the question of estimating the parameters using Bayesian uncertainty quantification. 
Due to a complex stochastic nature, the numerical model for platelet deposition does not have a tractable likelihood function. We use Approximate Bayesian Computation (ABC), a likelihood-free inference scheme, with an optimal application of HPC \citep{Dutta_2017_PASC} to provide a Bayesian way to estimate adhesion and aggregation rates given the deposition patterns observed in the Impact-R of platelets collected from a patient. 
Obviously, the
clinical applicability of the proposed technique to provide  a new 
platelet function test remains to be explored, but the numerical model~\citep{Chopard_2017} and the proposed inference scheme here, bring the technical
elements together to build a new class of medical tests.

In section~\ref{sec:bcground_model} we introduce the necessary background
knowledge about the platelet deposition model, whereas Section~\ref{sec:BI} recalls the concept of Bayesian inference and introduces the HPC framework of ABC used in this study. Then we illustrate the results of the parameter determination  for platelet deposition model using ABC methodology, collectively for 7 patients in Section~\ref{sec:results}. Clearly, the same methodology can be used to determine the parameter values for each individual patients in a similar manner for a CVD clinical test. Finally, in section~\ref{sec:conclusion} we conclude the paper and discuss its impact from a biomedical perspective.

\section{Background and Scientific Relevance}
\label{sec:bcground_model}

The Impact-R~\citep{shenkman:08} is a well-known platelet function
analyzer.  It is a cylindrical device filled in with whole blood from
a donor. Its lower end is a fixed disk, serving as a deposition
surface, on which platelets adhere and aggregate.  The upper end of
the Impact-R cylinder is a rotating cone, creating an adjustable shear
rate in the blood. Due to this shear rate, platelets move towards the
deposition surface, where they adhere or aggregate. Platelets
aggregate next to already deposited platelets, or on top of them, thus
forming clusters whose size increase with time. 
This deposition process has been successfully described with a
mathematical model in~\cite{Chopard_2015,Chopard_2017}.

The numerical model (coined $\Model$ in what follows)
requires five parameters that specify the deposition process and are
relevant for a bio-medical understanding of the phenomena. In short,
the blood sample in the Impact-R device contains an
initial number $\np(0)$ of non-activated platelets per $\mu\ell$ and a
number $\nacp(0)$ of pre-activated platelets per $\mu\ell$. Initially
both type of platelets are supposed to be uniformly distributed within
the blood. Due to the process known as shear-induced diffusion,
platelets hit the deposition surface. Upon such an event, an activated
platelets will adhere with a probability that depends on its adhesion
rate, $\Pad$, that we would like to determine. Platelets that have
adhered on the surface are the seed of a cluster that can grow due to
the aggregation of the other platelets reaching the deposition
surface. We denote with $\Pg$ the rate at which new platelets will
deposit next to an existing cluster. We also introduce $\Pt$ the rate
at which platelets deposit on top of an existing cluster. An important
observation made in~\cite{Chopard_2015,Chopard_2017} is that albumin,
which is abundant in blood, compete with platelet for deposition. This
observation is compatible with results reported in different
experimental settings~\citep{Remuzzi:93,Fontaine:2009,Sharma:81}.
As a consequence, the number of aggregation clusters and their size tends to
saturate as time goes on, even though there are still a large number
of platelets in suspension in the blood.

\begin{figure}
  \includegraphics[width=.35\textwidth]{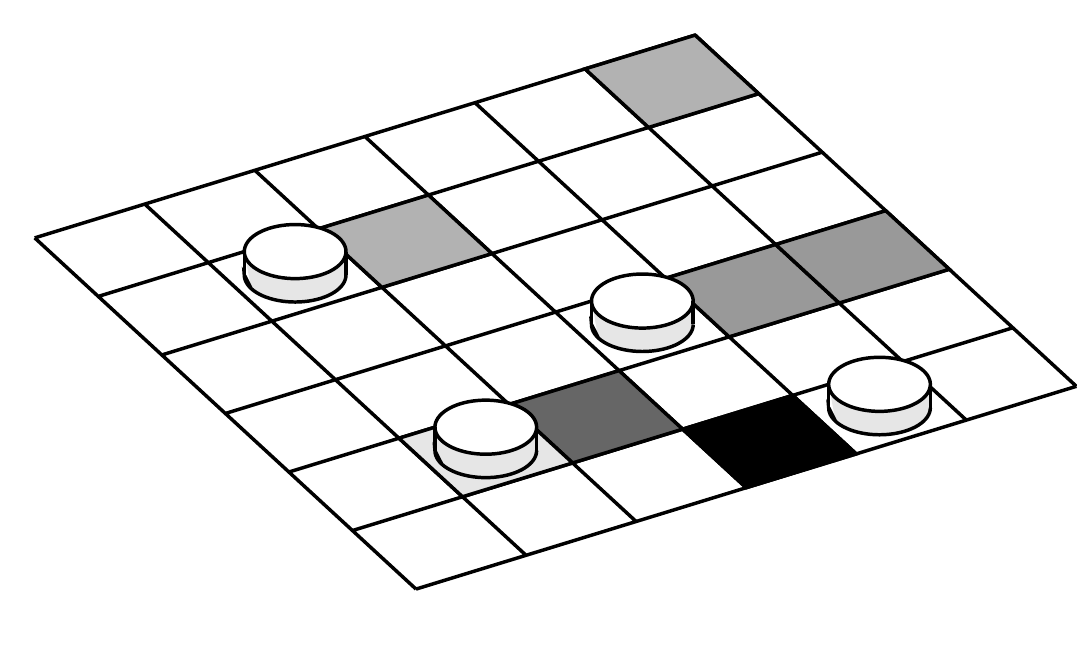}
  \hfill
  \includegraphics[width=.35\textwidth]{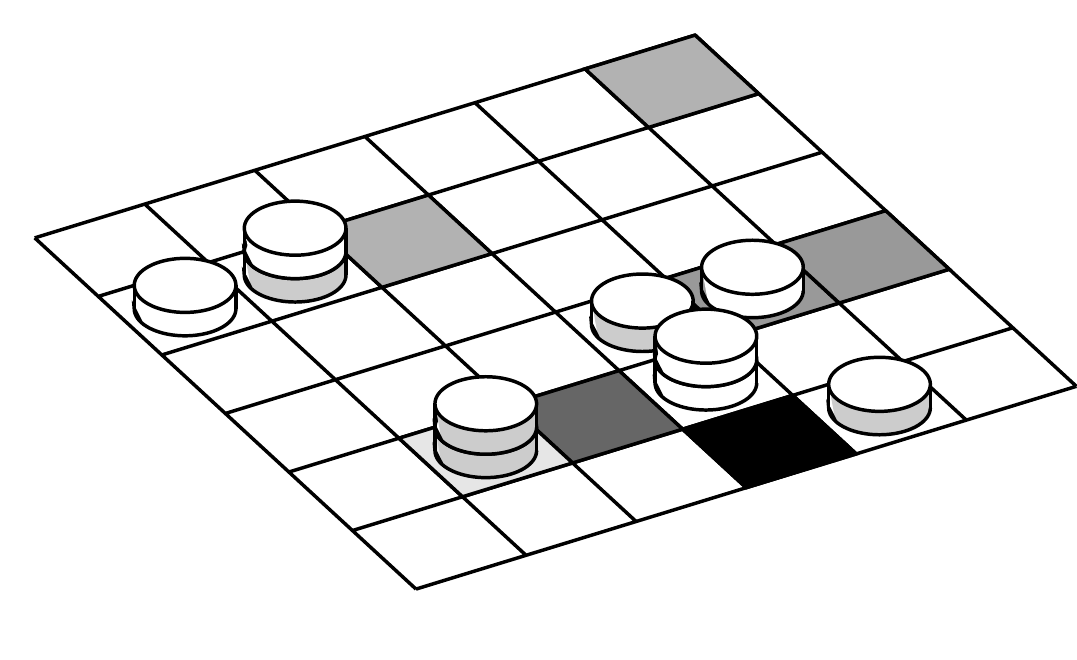}
  \caption{Sketch of the deposition substrate, discretized in cells of
    area equal to the surface of a platelet.  }
\label{fig:substrate}
 \end{figure}

To describe this process in
the model, two extra parameters, $\Pf$, the deposition
rate of albumin, and $\Ra$, a factor that accounts for the decrease
of platelets adhesion and aggregation on locations where albumin has
already deposited, were introduced. The numerical model is described in
  full detail in~\cite{Chopard_2015,Chopard_2017}. Here we simply
  repeat the main elements. Due to the mixing in the horizontal direction, it was assumed that the
activated platelets (AP), non-activated platelets (NAP) and albumin
(Al) in the bulk can be described by a 1D diffusion equation along the vertical axis $z$
\begin{equation}
  \partial_t\rho=D\partial^2_z\rho \qquad J=-D\grad\rho
  \label{eq:diffusion-1D}
\end{equation}
where $\rho$ is the density of either AP, NAP or Al, $J$ and $D$ are correspondingly the flux of
particles and the shear induced diffusion.
Upon reaching a boundary layer above the deposition substrate, adhesion and aggregation will take place according to
\begin{equation}
 \dot{N}=-J(0,t)\ds-p_{d}N(t)
\label{eq:N}
\end{equation}
where $N$ is the number of particles in the boundary layer, $\ds$ a
surface element on the deposition surface, and $p_{d}$ is the
deposition rate, which evolves during time and varies across the
substrate, according to the deposition history. For the deposition
process, particles are considered as discrete entities that can attach
to any position of the grid representing the deposition surface, as
sketched in Fig.~\ref{fig:substrate}. In this figure, the gray levels
illustrate the density of albumin already deposited in each cell. The
picture also illustrates the adhesion, aggregation, and vertical
deposition along the $z$-axis. On the left panel, activated platelets
(gray side disks) deposit first. Then in the second panel,
non-activated platelets (white side disks) aggregate next to an
already formed cluster. Both pre-activated and non-activated platelets can deposit on top of an existing cluster.

    The deposition rules are the following.  An albumin that reaches
    the substrate at time $t$ deposits with a probability $P(t)$ which
    depends on the local density $\rho_{al}(t)$ of already deposited
    Al. We assume that $P$ is proportional to the remaining free space
    in the cell,
\begin{equation}
  P(t)=\Pf(\rho_{max}-\rho_{al}(t)),
\end{equation}
where $\Pf$ is a parameter and $\rho_{max}$ is determined by the
constraint that at most 100,000 albumin particles can fit in a
deposition cell of area $\ds=5~(\mu m)^2$, corresponding to the size of
a deposited platelet (obtained as the smallest variation of cluster
area observed with the microscope).

An activated platelet that hits a platelet-free cell deposits
with a probability $Q$, where $Q$ decreases as the local concentration
$\rho_{al}$ of albumin increases. We assumed that
\begin{equation}
  Q=\Pad\exp(-\Ra\rho_{al}),
\end{equation}
where $\Pad$ and $\Ra$ are parameters. This expression can be
justified by the fact that a platelet needs more free space than an
albumin to attach to the substrate, due to their size difference.  In
other words, the probability of having enough space for a platelet,
decreases roughly exponentially with the density of albumin in the
substrate.  This can be validated with a simple deposition model on a
grid, where small and large objects compete for deposition.

Once an activated platelet has deposited, it is the seed of a new
cluster that grows further due to the aggregation of further
platelets. In our model, AP and NAP can deposit next to already
deposited platelets.  From the above discussion, the aggregation
probability $R$ is assumed to be
\begin{equation}
  R=\Pg\exp(-\Ra\rho_{al}),
\end{equation}
with $\Pg$ another parameter.

The above deposition probabilities can also be expressed as deposition
rate over the given simulation time step $\dt=0.01~s$
(see~\cite{Chopard_2017} for details), hence giving a way to couple the
diffusion equation~(\ref{eq:diffusion-1D}) with the 2D discrete
deposition process sketched in Fig.~\ref{fig:substrate}. Particles
that did not deposit at time $t$ are re-injected in the bulk and
contribute to boundary condition of eq.~(\ref{eq:diffusion-1D}) at
$z=0$.

 \begin{figure}
\centering
 \includegraphics[width=.35\textwidth]{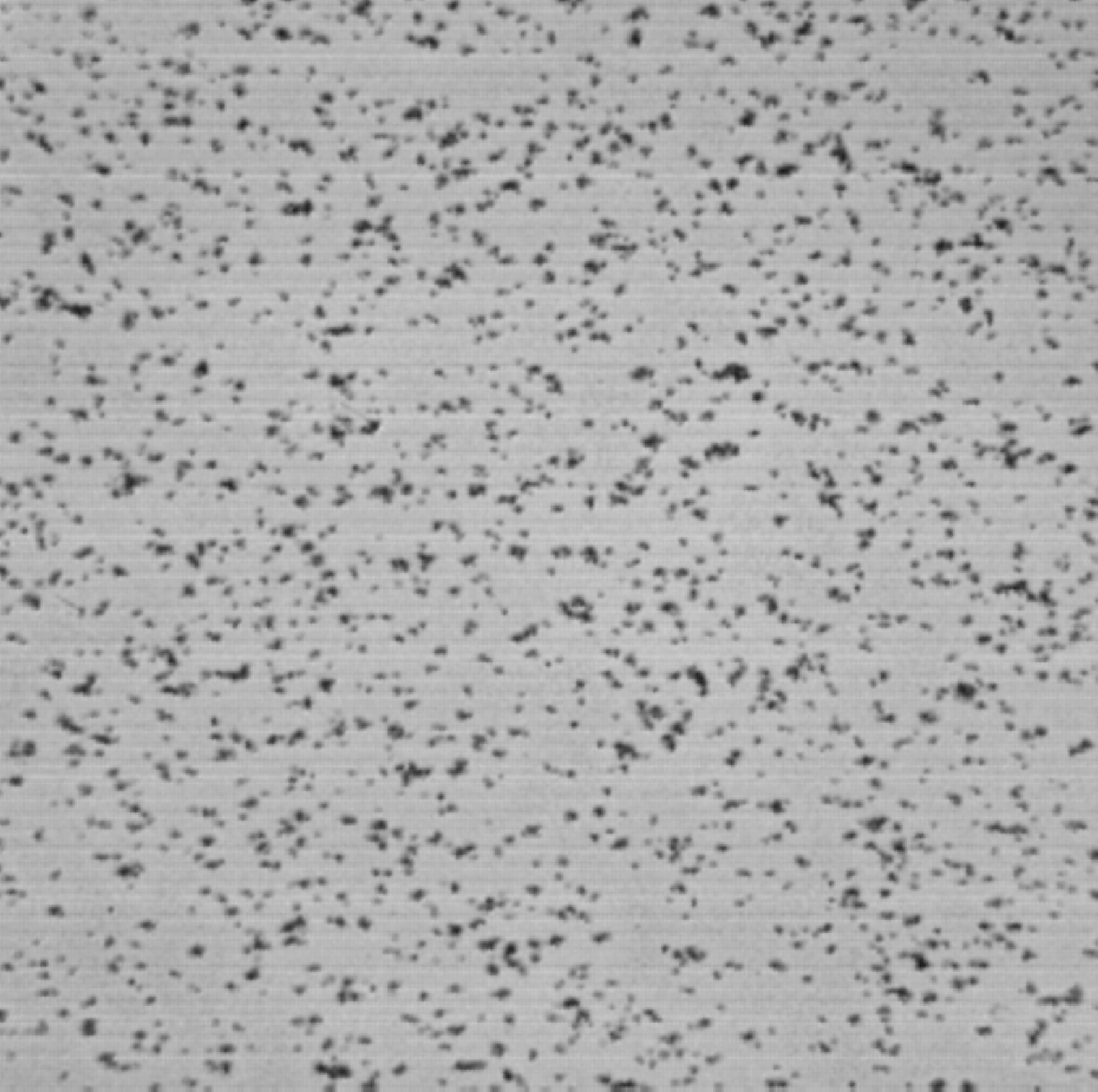}
 \hfill
 \includegraphics[width=.35\textwidth]{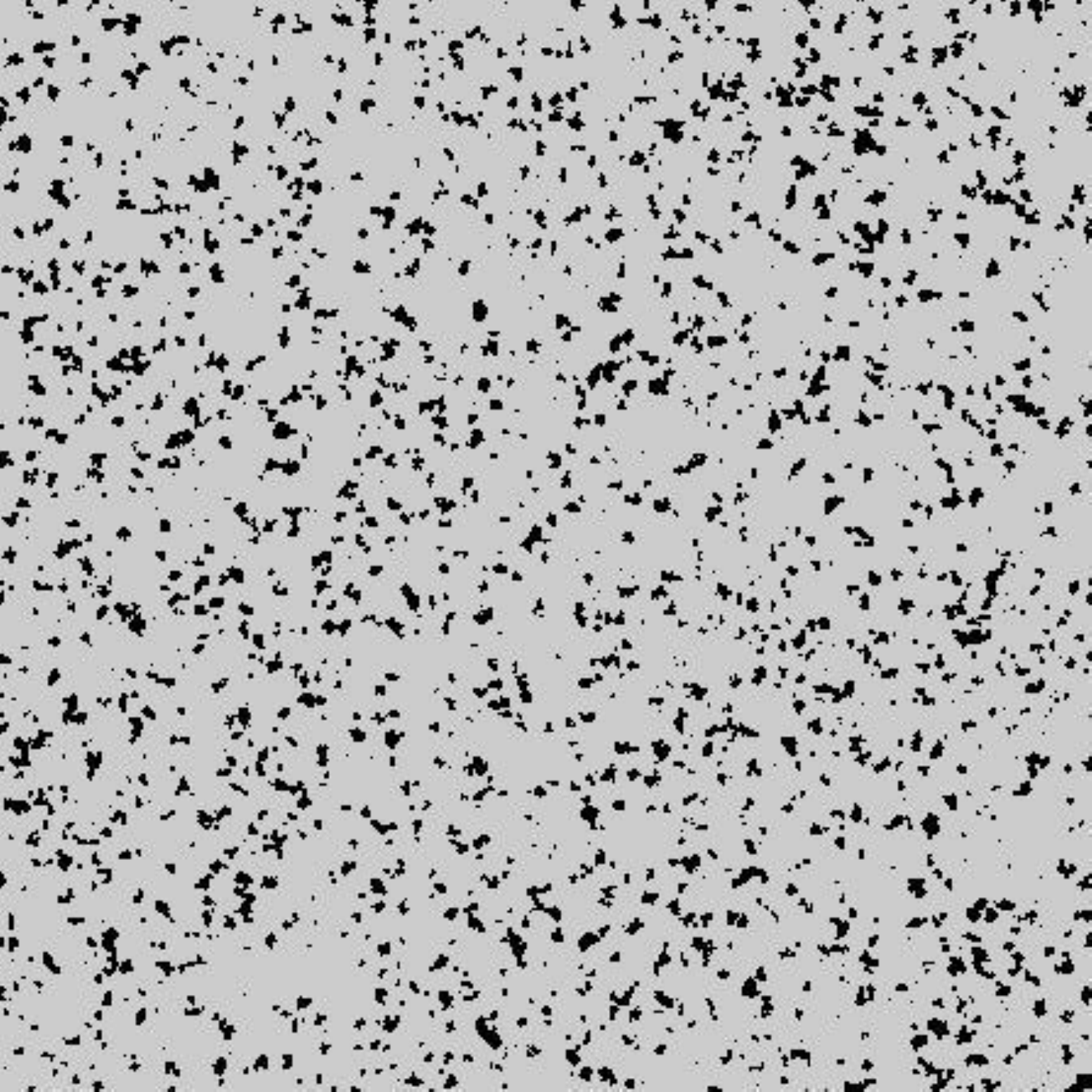}
\caption{The deposition surface of the Impact-R device after 300
  seconds (left) and the corresponding results of the deposition in
  the mathematical model (right). Black dots represent the deposited
  platelets that are grouped in clusters.}
\label{fig:impact-R}
\end{figure}

  To the best of our knowledge, except
  for~\cite{Chopard_2015,Chopard_2017} there is no model in the
  literature that describes quantitatively the proposed in-vitro
  experiment. The closest approach is that of \cite{affeld:13},
  but albumin is not included, and the role of pre-activated and
  non-activated platelets is not differentiated. Also, we are not
  aware of any other  study than ours that reports both the amount of platelets
  in suspension as a function of time and those on the deposition surface.

  The validity of the proposed numerical model has been explored in
  detail in~\cite{Chopard_2017}. This validation is based on the fact
  that the model, using hand-tuned parameters can reproduce the
  time-dependent experimental observations very well. We refer the
  readers to~\cite{Chopard_2017} for a complete discussion. Here we
  briefly recall the main elements that demonstrate the excellent
  agreement of the model and the simulations.  We reproduce
  Fig.~\ref{fig:impact-R} from \cite{Chopard_2017}, showing the visual
  similarity between the actual and simulated deposition pattern. In
  the validation study, the evolution of the number of clusters, their
  average size and the numbers of pre-activated and non-activated
  platelets still in suspension matched quantitatively with the
  experimental measurements at times 20, 60, 120 and 300~s. In
  addition, a very good agreement between the simulated deposition
  pattern and the experiment was also found by comparing the
  distributions of the areas and volumes of the aggregates.
 
To be noticed, the validation reported in \cite{Chopard_2017} was done using
manually estimated parameters. As the main goal of this research is to
propose an inference scheme to learn the parameters in a data-driven
manner, a validation for the model and the inference scheme is
reported in Fig.~\ref{fig:prediction_check_experimental} below, using
the inferred posterior distribution which also includes a
quantification of prediction error.
  
For the purpose of the present study, the model $\Model$ is
parametrized in terms of the five quantities introduced above, namely
the adhesion rate $\Pad$, the aggregation rates $\Pg$ and $\Pt$, the
deposition rate of albumin $\Pf$, and the attenuation factor $\Ra$. Some additional parameters of the model, specifically, the
shear-induced diffusion coefficient and the thickness of the boundary
layer \citep{Chopard_2017}, are assumed here to be
known.
Collectively, we define
\[ \parameter= (\Pg, \Pad, \Pt, \Pf, \Ra). \]
If the initial values for $\np(0)$ and $\nacp(0)$,
as well as the concentration of albumin are known from the experiment, we
can forward simulate the deposition of platelets over time using 
model $\Model$ for the given values of these parameters $\parameter = \parameter^*$:
\begin{eqnarray}
\label{eq:simulator_depo}
\Model [\parameter = \parameter^*] \rightarrow \left\lbrace 
\left(\sac(t),\nac(t),\np(t),\nacp(t)\right), \ t=0, \ldots, T \right\rbrace.
\end{eqnarray}
where $\sac(t),\nac(t),\np(t)$ \mbox{ and } $\nacp(t)$ are
correspondingly average size of the aggregation clusters, their number
per $mm^2$, the number of non-activated and pre-activated platelets
per $\mu\ell$ still in suspension at time $t$.

The Impact-R experiments have been repeated with the whole blood
obtained from seven donors and the observations were made at time, 0 sec., 20 sec.,
60 sec., 120 sec. and 300 sec. At these five time points, $\left(\sac(t), \nac(t),
\np(t), \nacp(t) \right)$ are measured. Let us call the observed dataset 
collected through experiment as,
\[ \dataObs \equiv \lbrace (\sac^0(t),\nac^0(t),\np^0(t)
,\nacp^0(t)): t = 0 \mbox{ sec.}, \ldots, 300 \mbox{ sec.} \rbrace.\] By comparing the number 
and size of the deposition aggregates obtained
from the {\it in-vitro} experiments with the computational results
obtained by forward simulation from the numerical model (see
Fig.~\ref{fig:impact-R} for an illustration), the model parameters were manually calibrated by a trial and error procedure in \cite{Chopard_2017}. Due to the complex nature of the model and high-dimensional parameter space, this manual determination of the parameter values are subjective and time consuming.   

However, if the parameters of the model could be learned
more rigorously with an automated data-driven methodology, we could
immensely improve the performance of these models and bring this scheme
 as a new clinical test for platelet functions. To this aim, here we propose to use ABC for Bayesian inference 
of the parameters. As a result of Bayesian inference
to this context, not only we can automatically and efficiently 
estimate the model parameters, but we
can also perform parameter uncertainty quantification in a
statistically sound manner, and determine if the provided solution is
unique. 

\section{Bayesian Inference}
\label{sec:BI}
We can quantify the  uncertainty of the unknown parameter $\parameter$ by a posterior distribution $p(\parameter|\data)$ given the observed dataset $\data = \dataObs$. A posterior distribution is obtained, by Bayes' Theorem as,
\begin{eqnarray}
p(\parameter|\data) = \frac{\prior(\parameter)p(\data|\parameter)}{m(\data)},
\end{eqnarray}
where $\prior(\parameter)$, $p(\data|\parameter)$ and $m(\data) = \int\prior(\parameter)p(\data|\parameter)d\parameter$ are correspondingly the prior distribution on the parameter $\parameter$, the likelihood function, and the marginal likelihood. The prior distribution $\prior(\parameter)$ ensures a way to leverage the learning of parameters with prior knowledge, which is commonly known due to the availability of medical knowledge regarding cardio-vascular diseases. If the likelihood function can be evaluated, at least up to a normalizing constant, then the posterior distribution can be approximated by drawing a sample of parameter values from the posterior distribution using (Markov chain) Monte Carlo sampling schemes \citep{Robert_2005}. For the simulator-based models considered in Section~\ref{sec:bcground_model}, the likelihood function is difficult to compute as 
it requires solving a very high dimensional integral. In next Subsection~\ref{sec:Inference_method}, we illustrate ABC to perform Bayesian Inference for models where the analytical form of the likelihood function is not available in closed form or not feasible to compute. 

\subsection{Approximate Bayesian computation} 
\label{sec:Inference_method}
ABC allows us to draw samples from the approximate posterior distribution of parameters of the simulator-based models in absence of likelihood function, hence to perform approximate statistical inference (eg., point estimation, hypothesis testing, model selection etc.) in a data-driven manner. In a fundamental Rejection ABC scheme, we simulate from the model $\Model(\parameter)$ a synthetic dataset $\datasim$ for a parameter value $\parameter$ and measure the closeness between $\datasim$ and $\dataObs$ using a pre-defined discrepancy function $\distance(\datasim,\dataObs)$. Based on this discrepancy measure, ABC accepts the parameter value $\parameter$ when $\distance(\datasim,\dataObs)$ is less than a pre-specified threshold value $\epsilon$. 

As the Rejection ABC scheme is computationally inefficient, to explore the parameter space in an efficient manner, there exists a large group of ABC algorithms \citep{Marin_2012}. 
As pointed in \cite{Dutta_2017_PASC}, these ABC algorithms, consist of four fundamental steps: 
\begin{itemize}
\item[1.] (Re-)sample a set of parameters $\parameter$ either from the prior distribution or from an already existing set of parameter samples; 
\item[2.] For each of the sample from the whole set or a subset, perturb it using the perturbation kernel, accept the perturbed sample based on a decision rule governed by a threshold or repeat the whole second step; 
\item[3.] For each parameter sample calculate its weight; 
\item[4.] Normalize the weights, calculate a co-variance matrix and adaptively re-compute the threshold for the decision rule.
\end{itemize}
 These four steps are repeated until the weighted set of parameters, interpreted as the approximate posterior distribution, is `sufficiently close' to the true posterior distribution. The steps (1) and (4) are usually quite fast, compared to steps (2) and (3), which are the computationally expensive parts. 

These ABC algorithms can be generally classified into two groups based on the decision rule in step (2). In the first group, we simulate $\datasim$ using the perturbed parameter and accept it if $d(\datasim, \dataObs)< \epsilon$, an adaptively chosen threshold. Otherwise we continue until we get an accepted perturbed parameter. For the second group of algorithms, we do not have this `explicit acceptance' step but rather a probabilistic one. Here we accept the perturbed parameter with a probability that depends on $\epsilon$; if it is not accepted, we keep the present value of the parameter. The algorithms belonging to the `explicit acceptance' group are RejectionABC~\citep{tavare:1997} and PMCABC~\citep{beaumont:2010}, whereas the algorithms in the `probabilistic acceptance' group are  SMCABC~\citep{del:2012}, RSMCABC~\citep{drovandi2011estimation}, APMCABC~\citep{lenormand2013adaptive}, SABC~\citep{Albert_2015} and ABCsubsim~\citep{chiachio2014approximate}.
For an `explicit acceptance' to occur, it may take different amounts of time for different perturbed parameters (more repeated steps are needed if the proposed parameter value is distant from the true parameter value). Hence the first group of algorithms are inherently imbalanced. We notice that an ABC algorithm with `probabilistic acceptance' do not have the similar issue of imbalance as a probabilistic acceptance step takes approximately the same amount of time for each parameter.

The generation of $\datasim$ from the model, for a given parameter value, usually takes up huge amounts of computational resources (e.g. 10 minutes for the platelets deposition model in this paper). 
Hence, we want to choose an algorithm with faster convergence to the posterior distribution with minimal number of required forward simulations. For this work we choose Simulated Annealing ABC (SABC) which uses a probabilistic decision rule in Step (2) and needs minimal number of forward simulation than other algorithms as shown in \cite{Albert_2015}. As all tasks of SABC in Step (2) can be run independently, in our recent work \citep{Dutta_2017_PASC}, we have adapted SABC for HPC environment. Our implementation is available in Python package ABCpy and shows a linear scalability.  

We further note that the parallelization schemes
in ABCpy were primarily meant for inferring parameters from models, for which forward simulation takes almost equal time for any values of $\parameter$. Due to the complex stochastic nature of
the numerical model, forward simulation time for different values of $\parameter$, can be quite variable. To solve this imbalance in the forward simulation, additionally to the imbalance reported for ABC algorithms, we use a new dynamic allocation scheme for MPI developed in \cite{dutta2017abcpyhpc}.

\begin{figure}[h]
  \centering
\includegraphics[width=0.45\textwidth]{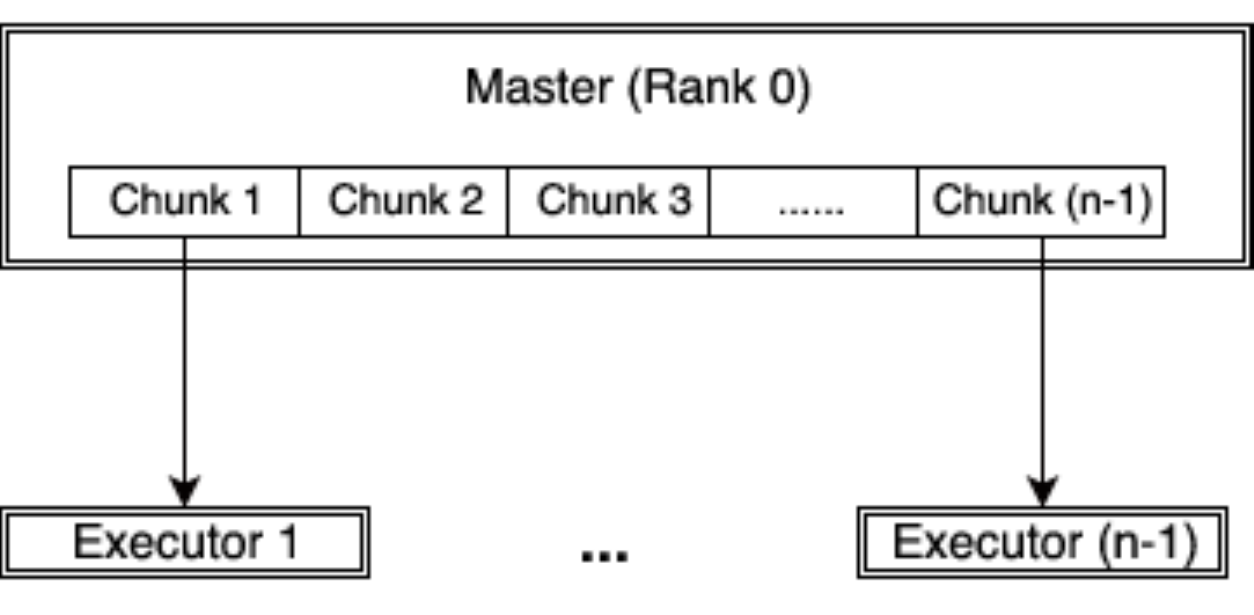}\label{fig:MPI_workflow_basic}
  \hfill
\includegraphics[width=0.45\textwidth]{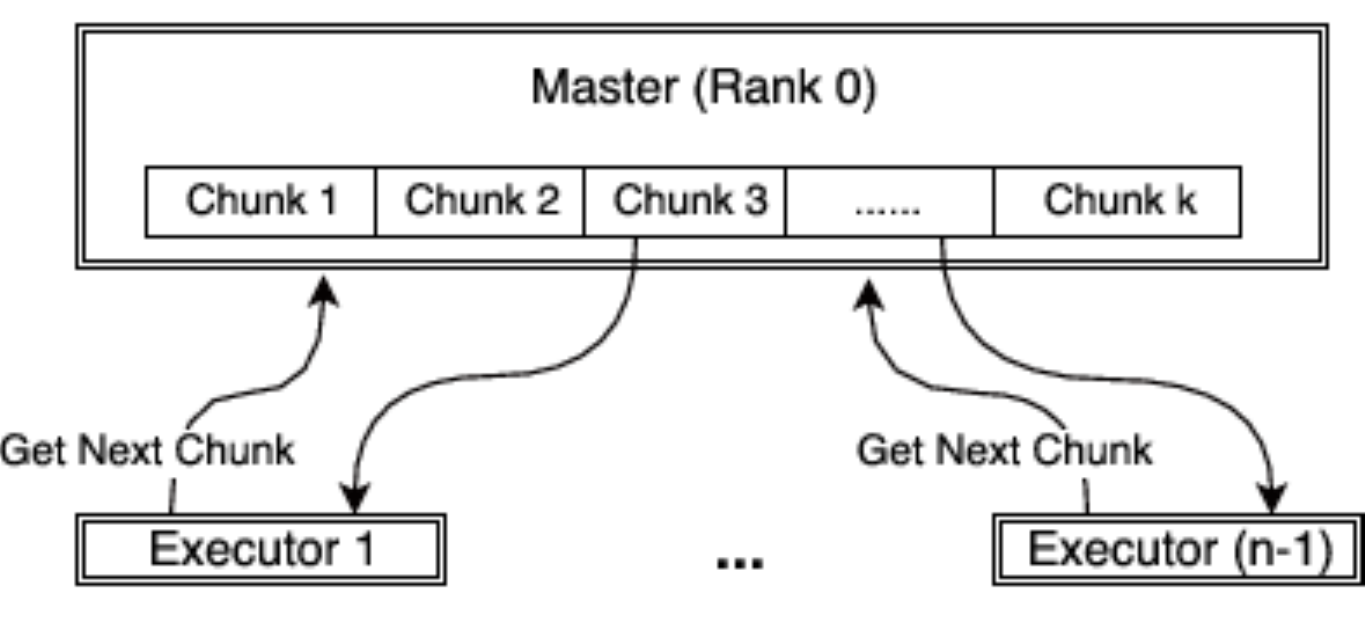}\label{fig:MPI_workflow_dynamic}
   \hfill
  \caption{Comparison of work-flow between MPI \textbf{(left)} and dynamic-MPI backend \textbf{(right)}.}
  \label{fig:MPI_workflow_comparison}
\end{figure}

\begin{figure}[h]
  \centering
\includegraphics[width=0.45\textwidth]{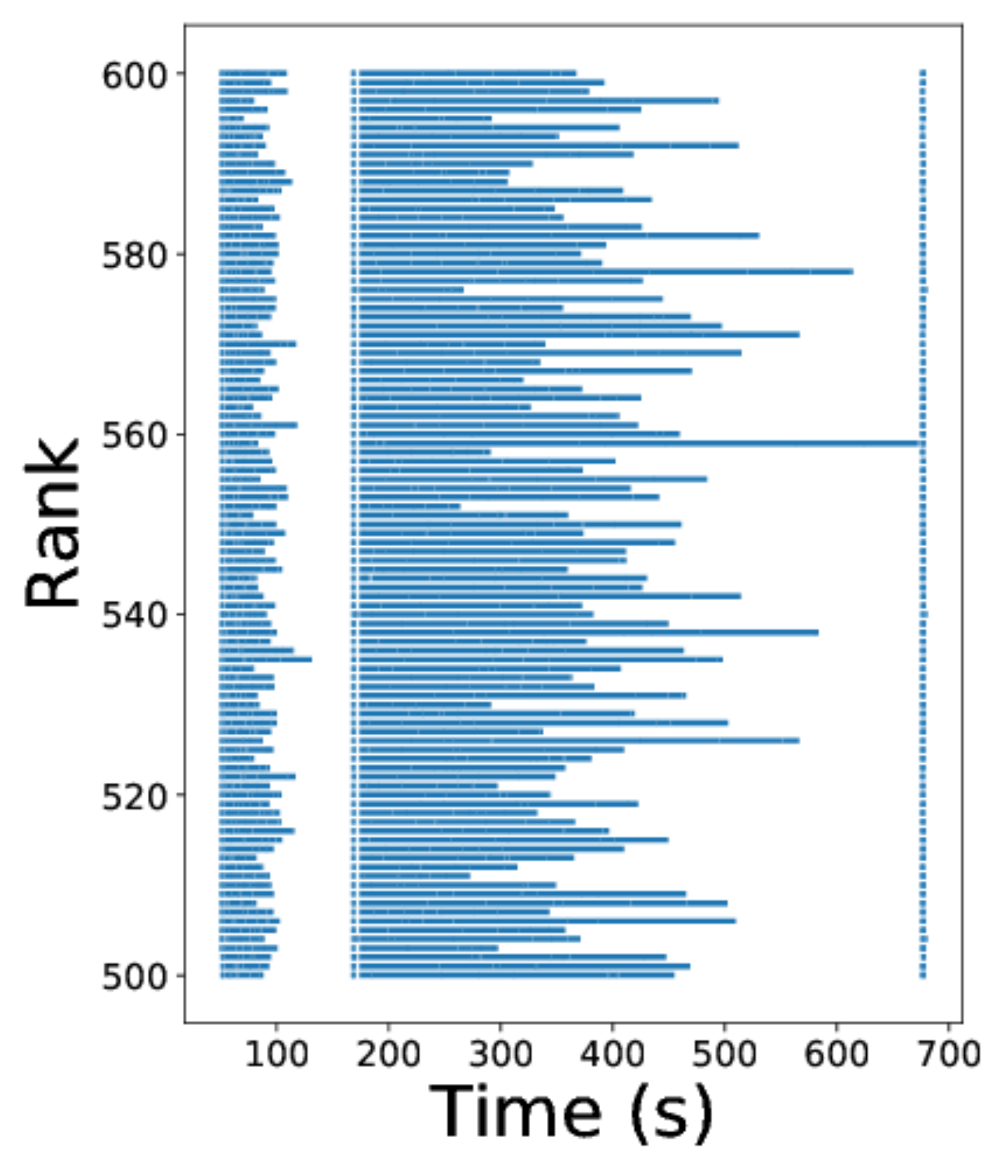}\label{fig:imbalance_apmcabc_mpi}
    \hfill
\includegraphics[width=0.45\textwidth]{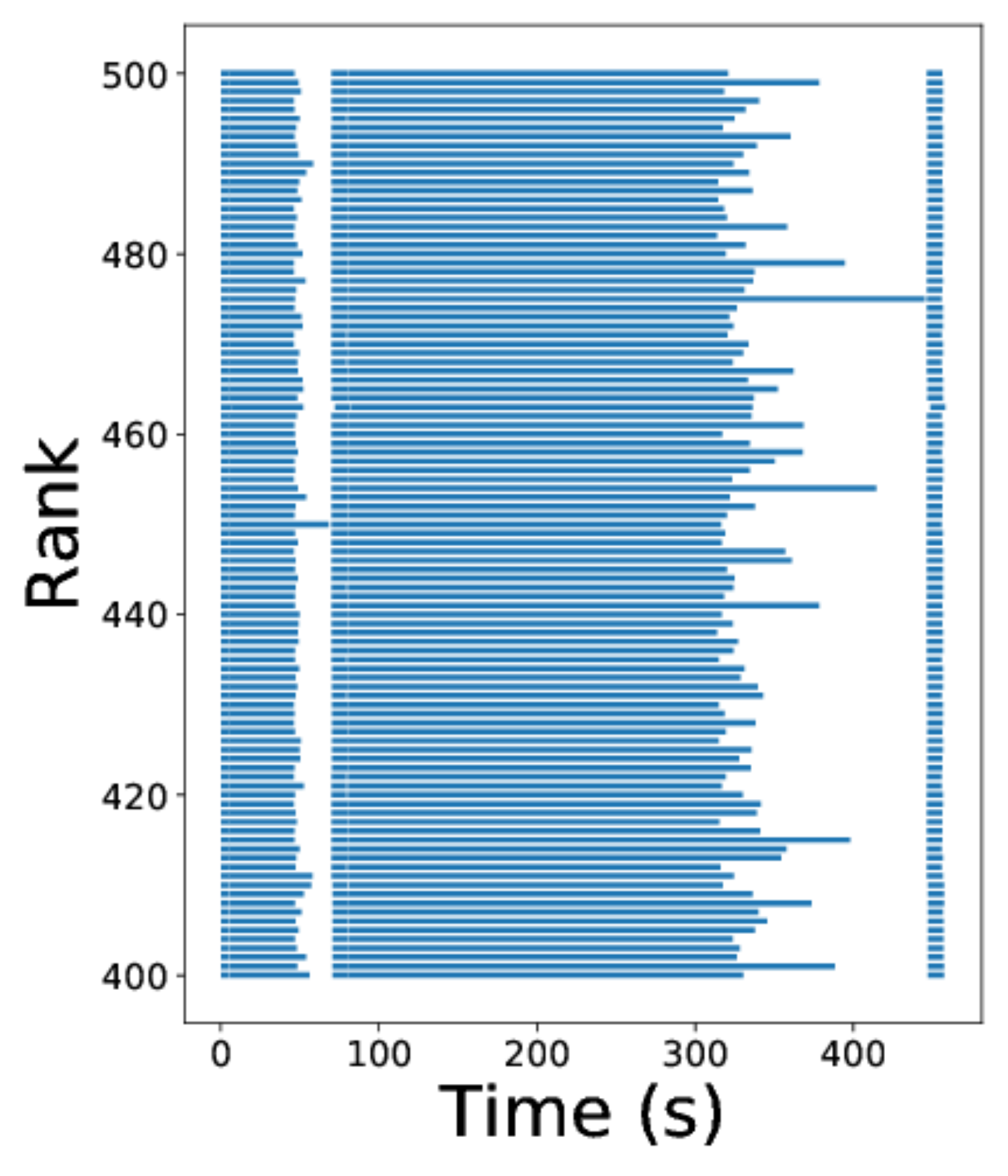}\label{fig:imbalance_apmcabc_dyn}
  \caption{Imbalance of ABC algorithms using MPI(straight-forward) \textbf{(left)} and MPI(dynamic-allocation) backend \textbf{(right)}.}
  \label{fig:imbalance_apmcabc}
\end{figure}

\subsection{Dynamic Allocation for MPI}
\label{sec:dynamicMPI}
Here we briefly discuss how a dynamic allocation strategy for map-reduce provides better balancing of ABC algorithms compared to a straightforward allocation approach. 

In the straightforward approach, the allocation scheme initially distributes $m$ tasks to $n$ executors, sends the map function to each executor, which in turn applies the map 
function, one after the other, to its $m/n$ map tasks. This approach is visualized in Figure~\ref{fig:MPI_workflow_comparison}, where a chunk represents the set of $m/n$ map tasks. For example, if we want to draw $10,000$ samples from the posterior distribution and we have $n = 100$ cores available, at each step of SABC we create groups of 100 parameters and each group is assigned to one individual core.

On the other hand, the dynamic allocation scheme initially distributes $k < m$ tasks to the $k$ executors, sends the map function to each executor, which in turn applies it to the
single task available. In contrast to the straightforward allocation, the executor requests a new map task as soon as the old one is terminated.
This clearly results in a better balance of the work.
The dynamic allocation strategy is an implementation of the famous
greedy algorithm for job-shop scheduling, 
which can be shown to have an overall processing time (makespan) up to twice as better than the best makespan \citep{Graham1966}. 

This approach is illustrated
in Figure~\ref{fig:MPI_workflow_comparison}, reproduced from \cite{dutta2017abcpyhpc}.
The unbalanced behavior is apparent if we visualize the run time of the individual map tasks on each executor. In Figure~\ref{fig:imbalance_apmcabc}, the individual map tasks processing time is shown for an ABC algorithm performing inference on a weather prediction model, reported in \cite{dutta2017abcpyhpc}. Each row corresponds to an executor (or rank) and each bar corresponds to the total time spent on all tasks assigned to the respective rank (row) for one map call. For the straightforward allocation strategy, one can easily verify that most of the ranks finish their map tasks in half the time of the slowest rank. This clearly leads to large inefficiencies. Conversely, using the dynamic allocation strategy, the work is more evenly distributed across the ranks. The unbalancedness is not a problem that can be overcome easily by adding resources, rather speed-up and efficiency can drop drastically compared to the dynamic allocation strategy with increasing number of executors. For a detailed description and comparison, we direct readers to \cite{dutta2017abcpyhpc}.

\subsection{Posterior Inference}
\label{sec:postinf}
Using SABC within HPC framework implemented in ABCpy \citep{Dutta_2017_PASC}, we draw $Z=5000$ samples approximating the posterior distribution $p(\parameter|\dataObs)$, while keeping all the tuning parameters for the SABC fixed at the default values suggested in ABCpy package, except the number of steps and the acceptance rate cutoff, which was chosen respectively as 30 and $1e^{-4}$. 
The parallelized SABC algorithm, using HPC makes it possible to perform the computation in 5 hours (using 140 nodes with 36-core of Piz Daint Cray
architecture (Intel Broadwell + NVidia TESLA P100)), which would have been impossible by a sequential algorithm. 
To perform SABC for the platelets deposition model, the summary statistics extracted from the dataset, discrepancy measure between the summary statistics, prior distribution of parameters and perturbation Kernel to explore the parameter space for inference are described next. 

\paragraph*{Summary statistics} Given a dataset, $\data \equiv \lbrace (\sac(t),\nac(t),\np(t),\nacp(t)):t = 0~\mbox{sec.},\ldots,300~\mbox{sec.} \rbrace$, we compute an array of summary statistics. 
\begin{eqnarray*}
\mathcal{F}: \ \data \rightarrow (\pmb{\mean}, \pmb{\var}, \pmb{\autcor}, \pmb{\cor}, \pmb{\crosscor})
\end{eqnarray*}
defined as following,
\begin{itemize}
\item[-] $\pmb{\mean} = (\mean_1,\mean_2,\mean_3,\mean_4)$, mean over time.
\item[-] $\pmb{\var} = (\var_1,\var_2,\var_3,\var_4)$, variance over time.
\item[-] $\pmb{\autcor} = (\autcor_1,\autcor_2,\autcor_3,\autcor_4)$, auto-correlation with lag 1. 
\item[-] $\pmb{\cor} = (\cor_1,\cor_2,\cor_3,\cor_4,\cor_5,\cor_6)$, correlation between different pairs of variables over time.
\item[-] $\pmb{\crosscor} = (\crosscor_1,\crosscor_2,\crosscor_3,\crosscor_4,\crosscor_5,\crosscor_6)$, cross-correlation with lag 1 between different pairs of variables over time.
\end{itemize}
The summary statistics, described above, are chosen to capture the \textit{mean} values, \textit{variances} and the \textit{intra-} and \textit{inter-} dependence of different variables of the time-series over time.   

\paragraph*{Discrepancy measure:} Assuming the above summary statistics contain the most \emph{essential} information about the likelihood function of the simulator-based model, we compute Bhattacharya-coefficient \citep{Bhattacharya_1943} for each of the variables present in the time-series using their mean and variance and Euclidean distances between different \textit{inter-} and \textit{intra-} correlations computed over time. Finally we take a mean of these discrepancies, such that, in the final discrepancy measure discrepancy between each of the summaries are equally weighted. The discrepancy measure between two datasets, $\data^1$ and $\data^2$ can be specified as, 
\begin{eqnarray*}
\label{eq:discrep_measure}
d(\data^1, \data^2) &\equiv& d(\mathcal{F}(\data^1),\mathcal{F}(\data^2))\\
 &=& \frac{1}{8}\sum_{i=1}^{4}(1-\exp(-\bc(\mean_i^1,\mean_i^2,\var_i^1,\var_i^2))) \\
 &+&\frac{1}{2}\sqrt{\frac{1}{16}\left(
 \sum_{i=1}^{4}(\autcor_i^1-\autcor_i^2)^2+ \sum_{i=1}^{6}(\cor_i^1-\cor_i^2)^2+ \sum_{i=1}^{6}(\crosscor_i^1-\crosscor_i^2)^2
 \right)},
\end{eqnarray*}
where $\bc(\mean^1,\mean^2,\var^1,\var^2) = \frac{1}{4}\log\left(\frac{1}{4}\left(\frac{\var^1}{\var^2}+\frac{\var^2}{\var^1}+2\right)\right)+\frac{1}{4}\left(\frac{(\mean^1-\mean^2)^2}{\var^1+\var^2}\right)$ is the Bhattacharya-coefficient \citep{Bhattacharya_1943} and $0 \leq \exp(-\bc(\bullet)) \leq 1$. Further, we notice the value of the discrepancy measure is always bounded in the closed interval $[0, 1]$. 

\paragraph*{Prior:} We consider independent Uniform distributions for the parameters with a pre-specified range for each of them, $\Pg \sim U(5,20)$, $\Pad \sim U(50,150)$, $\Pt \sim U(0.5e-3,3e-3)$, $\Pf \sim U(.1,1.5)$ and $\Ra \sim U(0,10)$.

\paragraph*{Perturbation Kernel:} To explore the parameter space of $\parameter = (\Pg, \Pad, \Pt, \Pf, \Ra) \in [5,20]\times[50,150]\times[0.5e-3,3e-3]\times[.1,1.5]\times[0,10]$, we consider a five-dimensional truncated multivariate Gaussian distribution as 
the perturbation kernel. SABC inference scheme
centers the perturbation kernel at the sample it is perturbing and updates the variance-covariance matrix of
the perturbation kernel based on the samples learned
from the previous step.

\subsection{Parameter estimation}
Given experimentally collected platelet deposition dataset $\dataObs$, our main interest is to estimate a value for $\parameter$. In decision theory, Bayes estimator minimizes posterior expected loss, $E_{p(\parameter|\dataObs)}(\lossfunc(\parameter,\bullet)|\dataObs)$ for an already chosen loss-function $\lossfunc$. If we have $Z$ samples $(\parameter_{i})_{i=1}^{Z}$ from the posterior distribution $p(\parameter|\dataObs)$, the Bayes estimator can be approximated as,
\begin{eqnarray}
\label{eq:Bayes_estimate}
\estparameter= \argmin_{\parameter} \frac{1}{M}\sum_{i=1}^M \lossfunc(\parameter_{i},\parameter). 
\end{eqnarray}
As we consider the Euclidean loss-function $\lossfunc(\parameter,\hat{\parameter}) = (\parameter-\hat{\parameter})^2$ as the loss-function, the approximate Bayes-estimator can be shown to be $\estparameter=E_{p(\parameter|\dataObs)}(\parameter) \approx \frac{1}{Z}\sum_{i=1}^Z \parameter_{i}$.

\section{Inference on experimental dataset}
\label{sec:results}
\begin{figure}[htbp]
\centering
 \includegraphics[width=.45\textwidth]{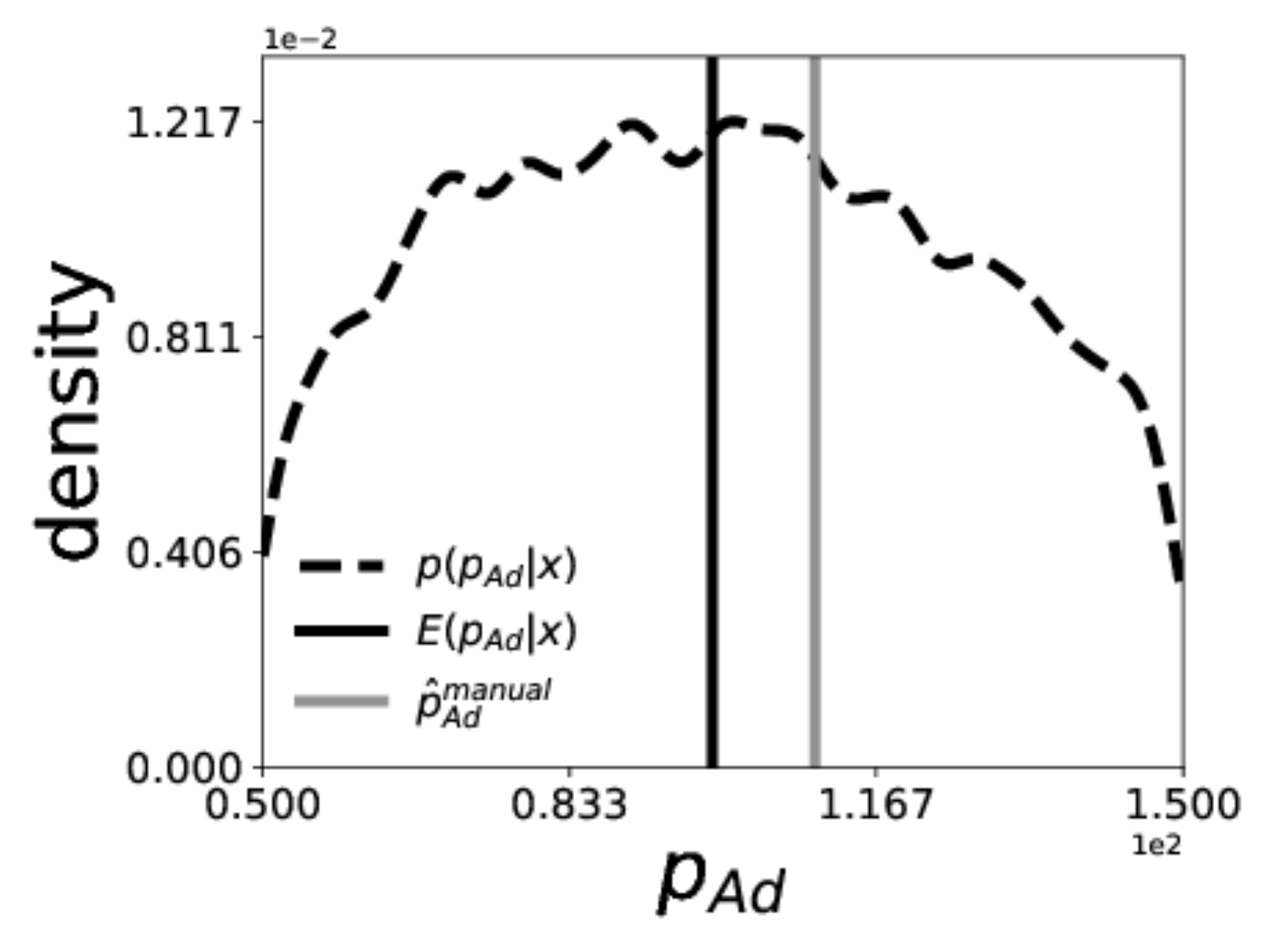}
 \hfill
 \includegraphics[width=.45\textwidth]{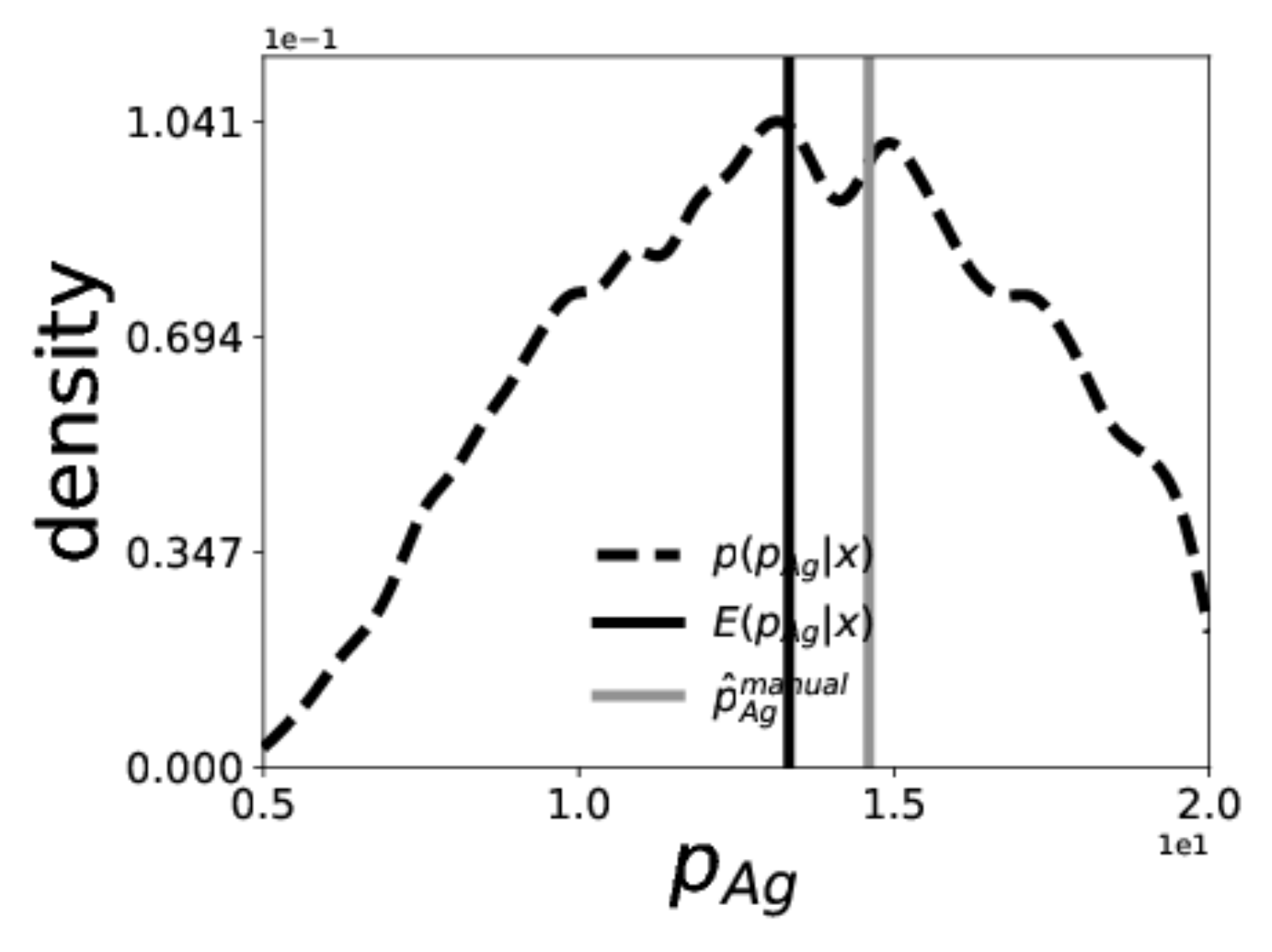}
 \hfill
 \includegraphics[width=.45\textwidth]{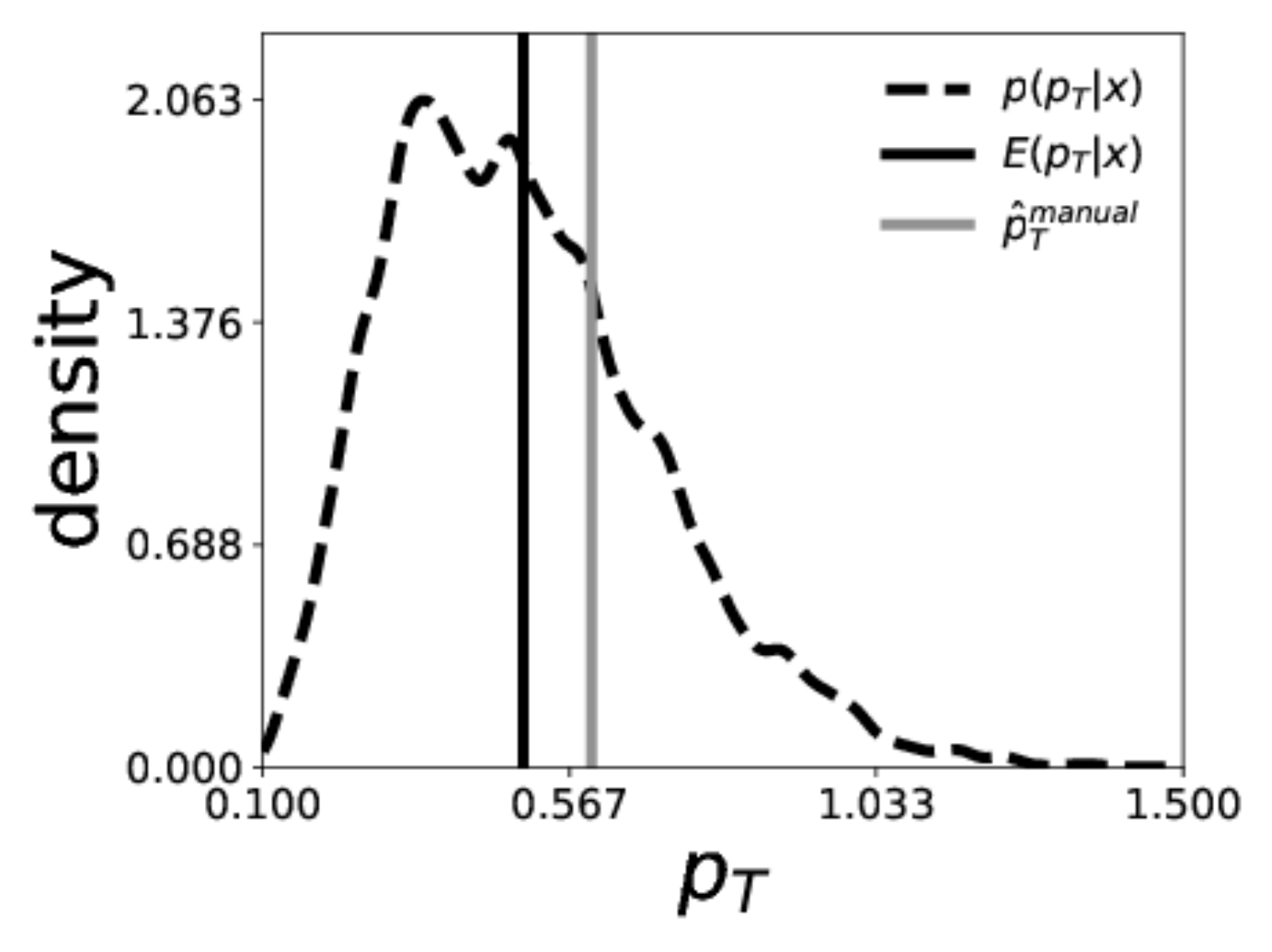}
 \hfill
 \includegraphics[width=.45\textwidth]{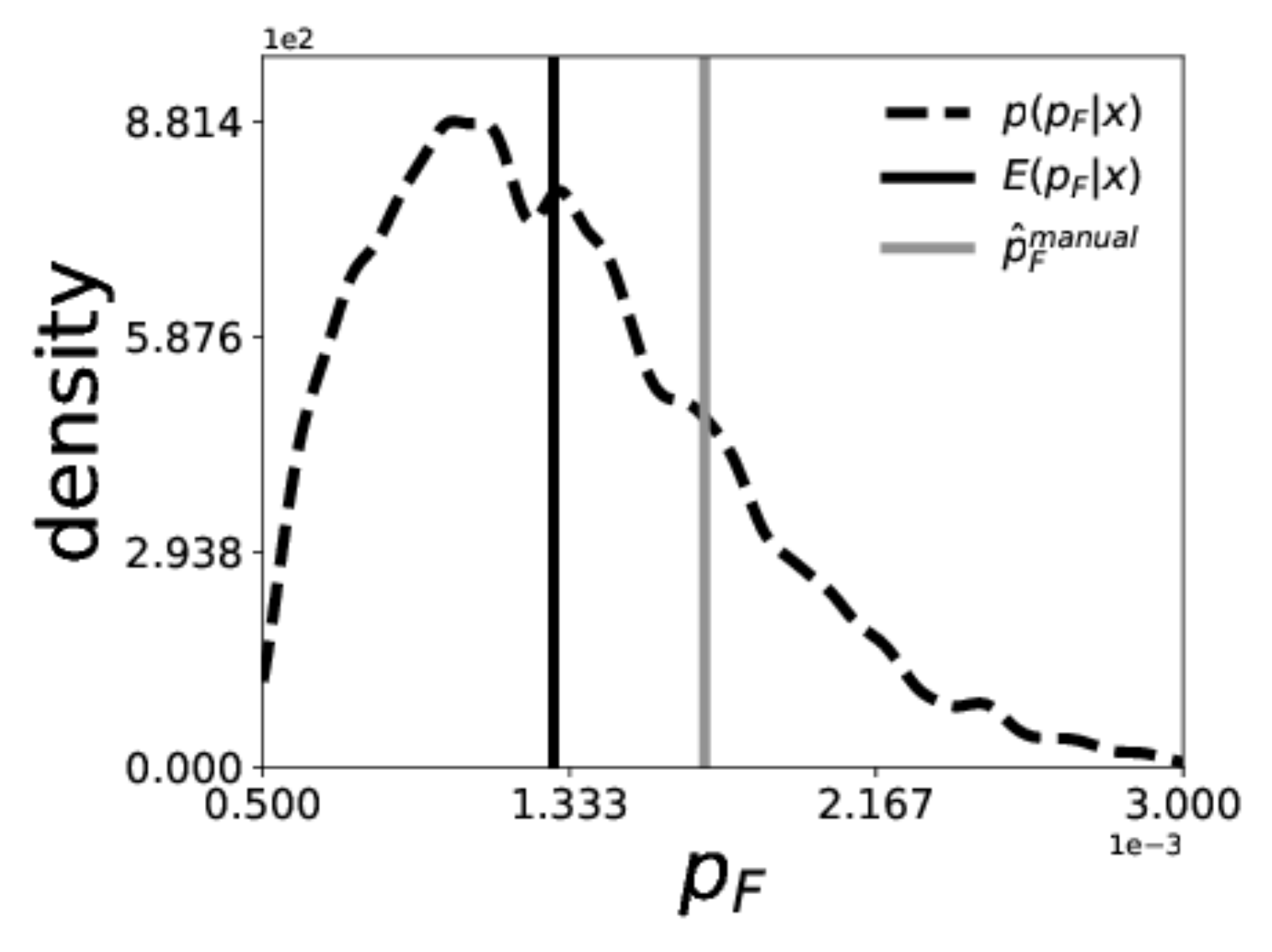}
 \hfill
 \includegraphics[width=.45\textwidth]{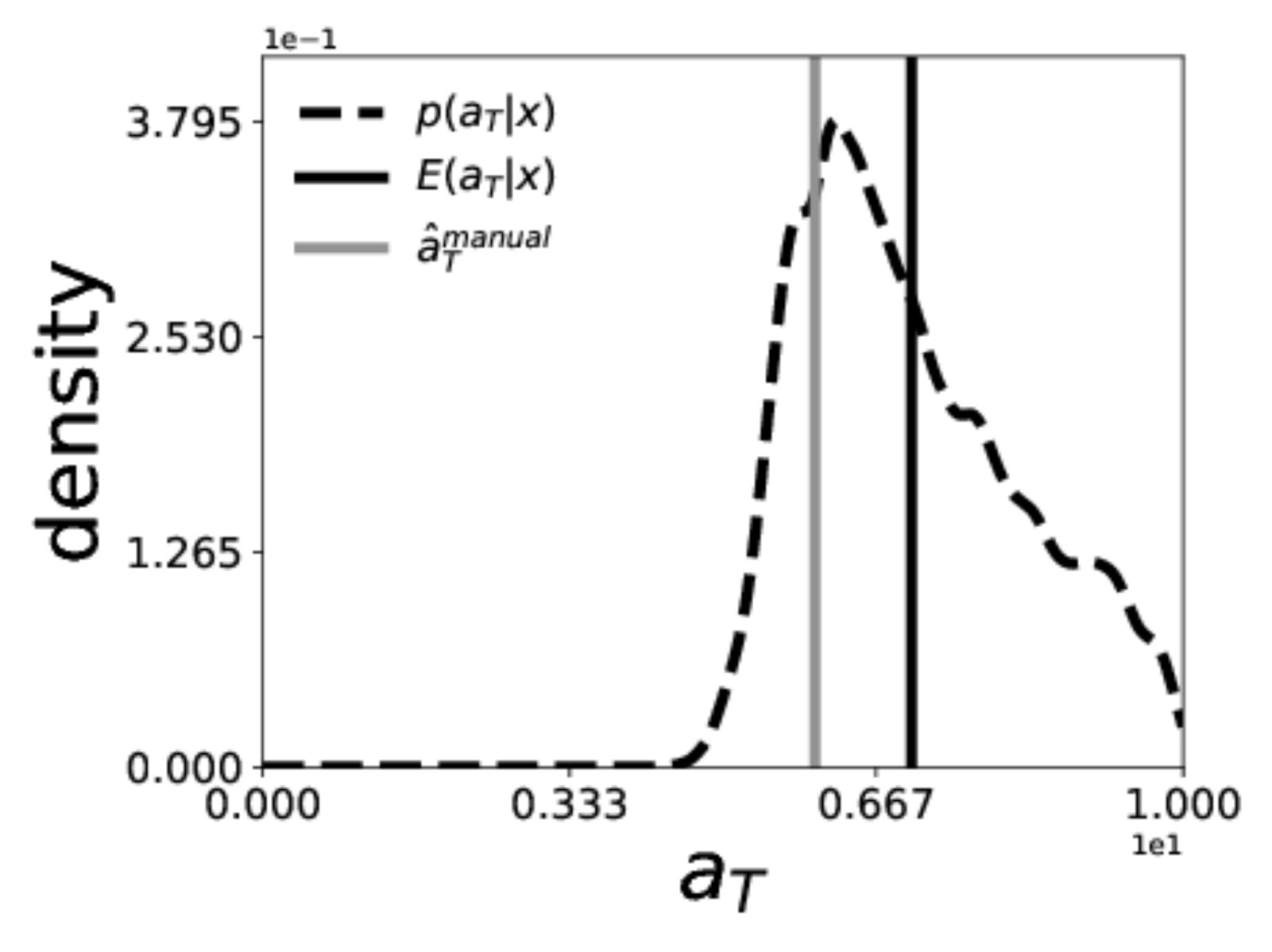}
\caption{Marginal posterior distribution (black-dashed) and Bayes Estimate (back-solid) of $\left(\Pad, \Pg, \Pt, \Pf, \Ra  \right)$ for collective dataset generated from of 7 patients. The smoothed marginal distribution is created by a Gaussian-kernel density estimator on 5000 i.i.d. samples drawn from the posterior distribution using SABC. The (gray-solid) line indicates the manually estimated values of the parameters in~\cite{Chopard_2017}.}
\label{fig:experimental_posterior_example}
\end{figure}

The performance of the inference scheme described in Section~\ref{sec:BI} is reported here, for a collective dataset created from the experimental study of platelets deposition of 7 blood-donors. The collective dataset was created by a simple average of $$\left(\sac(t), \nac(t), \np(t), \nacp(t) \right)$$ over 7 donors at each time-point $t$. In Figure~\ref{fig:experimental_posterior_example}, we show the Bayes estimate (black-solid) and the marginal posterior distribution (black-dashed) of each of the five parameters computed using 5000 samples drawn from the posterior distribution $p(\parameter|\dataObs)$ using SABC. For comparison, we also plot the manually estimated values of the parameters (gray-solid) in \cite{Chopard_2017}. We notice that the Bayes estimates are in a close proximity of the manually estimated values of the parameters and also the manually estimated values observe a significantly high posterior probability. This shows that, through the means of ABC we can get an estimate or quantify uncertainty of the parameters in platelets deposition model which is as good as the manually estimated ones, if not better. 

\begin{figure}[htbp]
\centering
 \includegraphics[width=.45\textwidth]{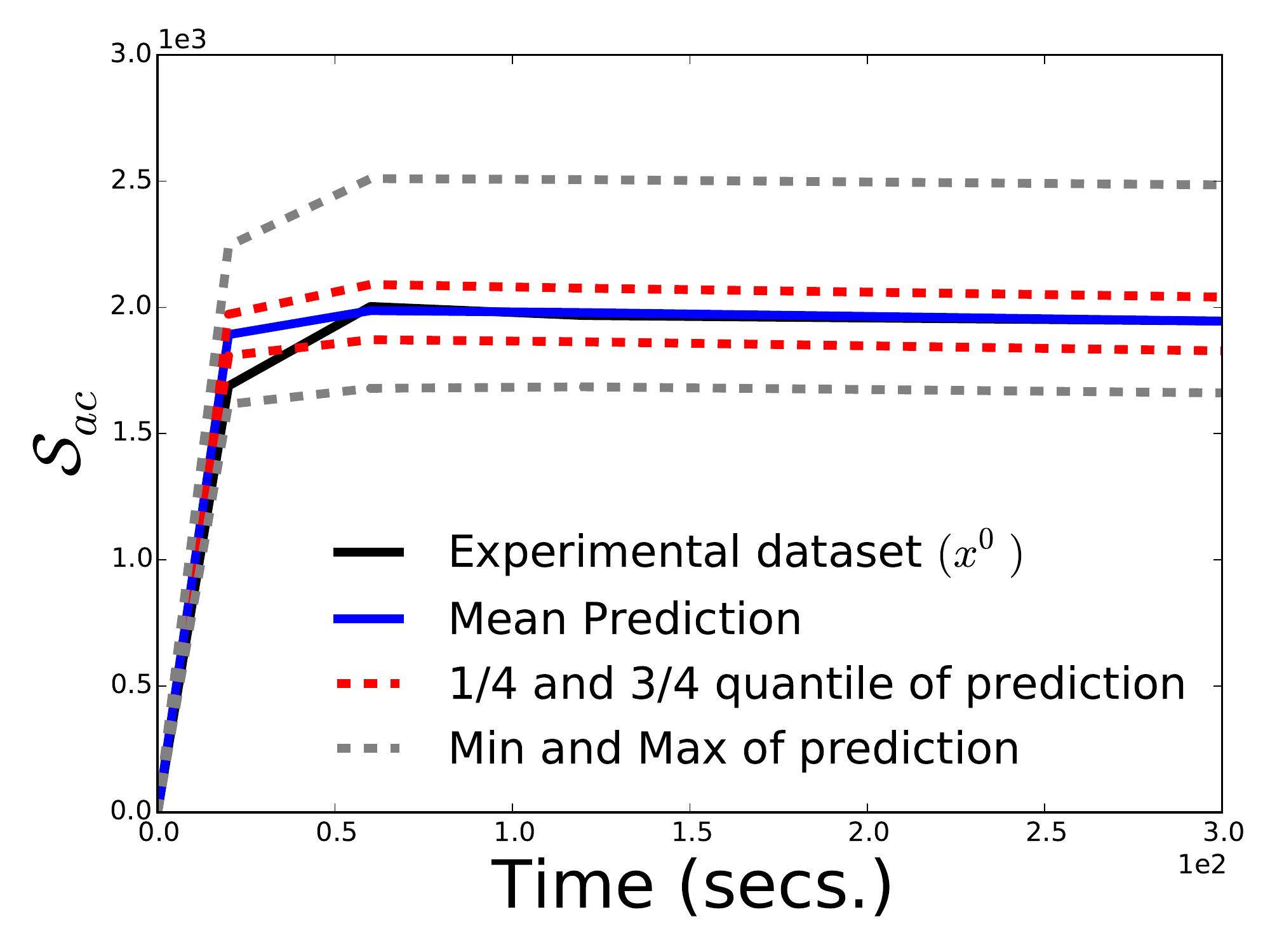}
 \hfill
 \includegraphics[width=.45\textwidth]{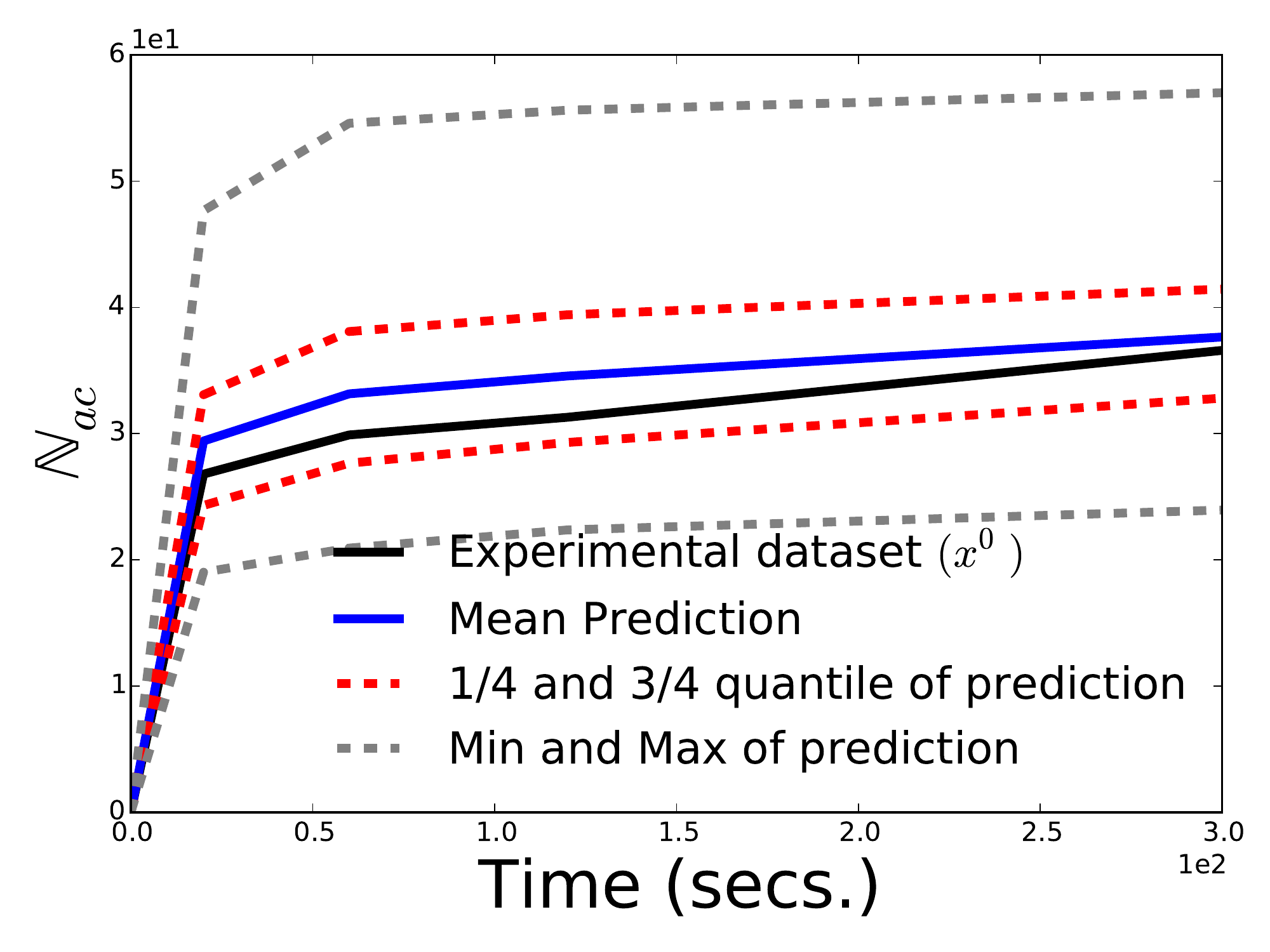}
 \hfill
 \includegraphics[width=.45\textwidth]{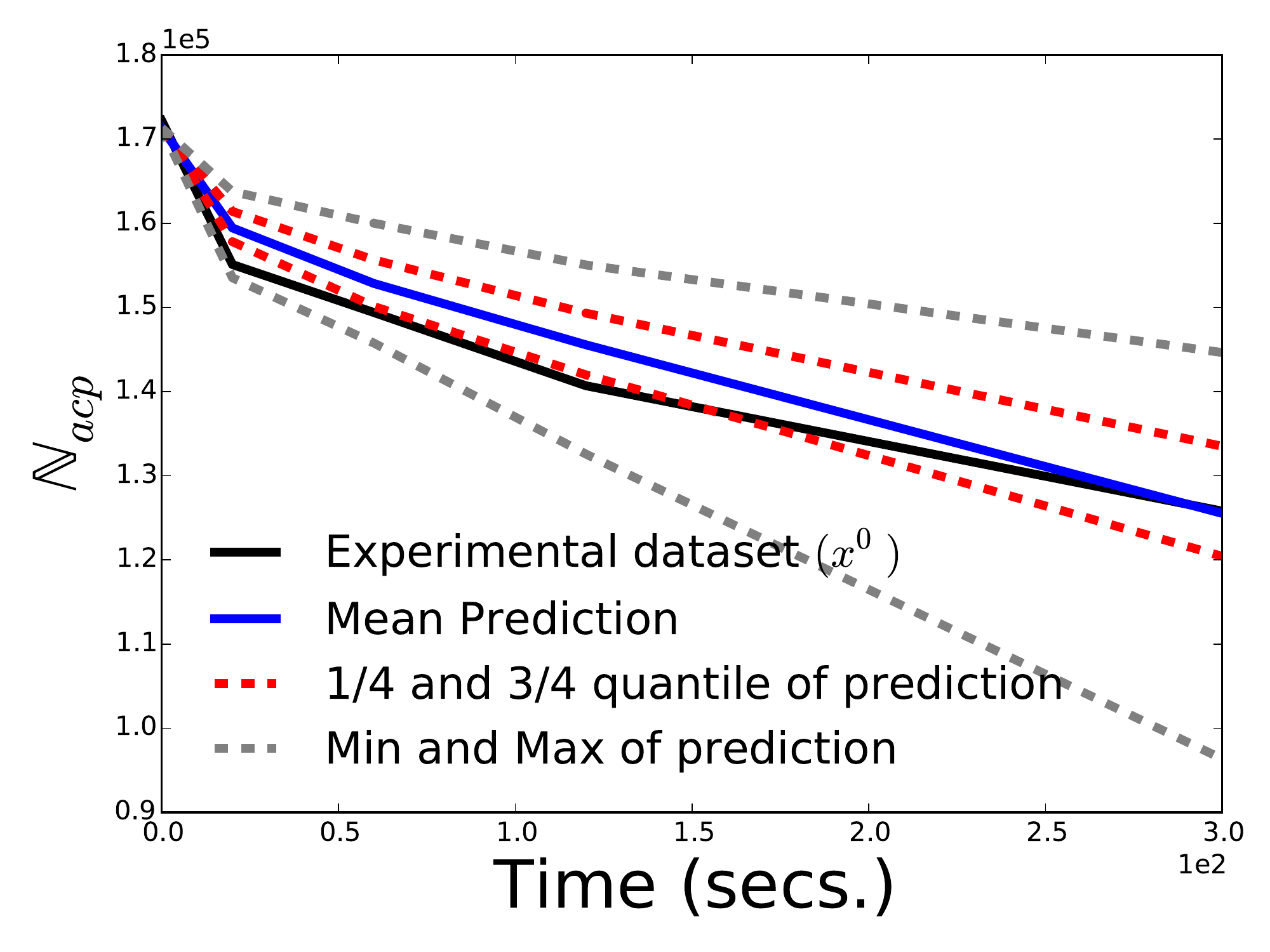}
 \hfill
 \includegraphics[width=.45\textwidth]{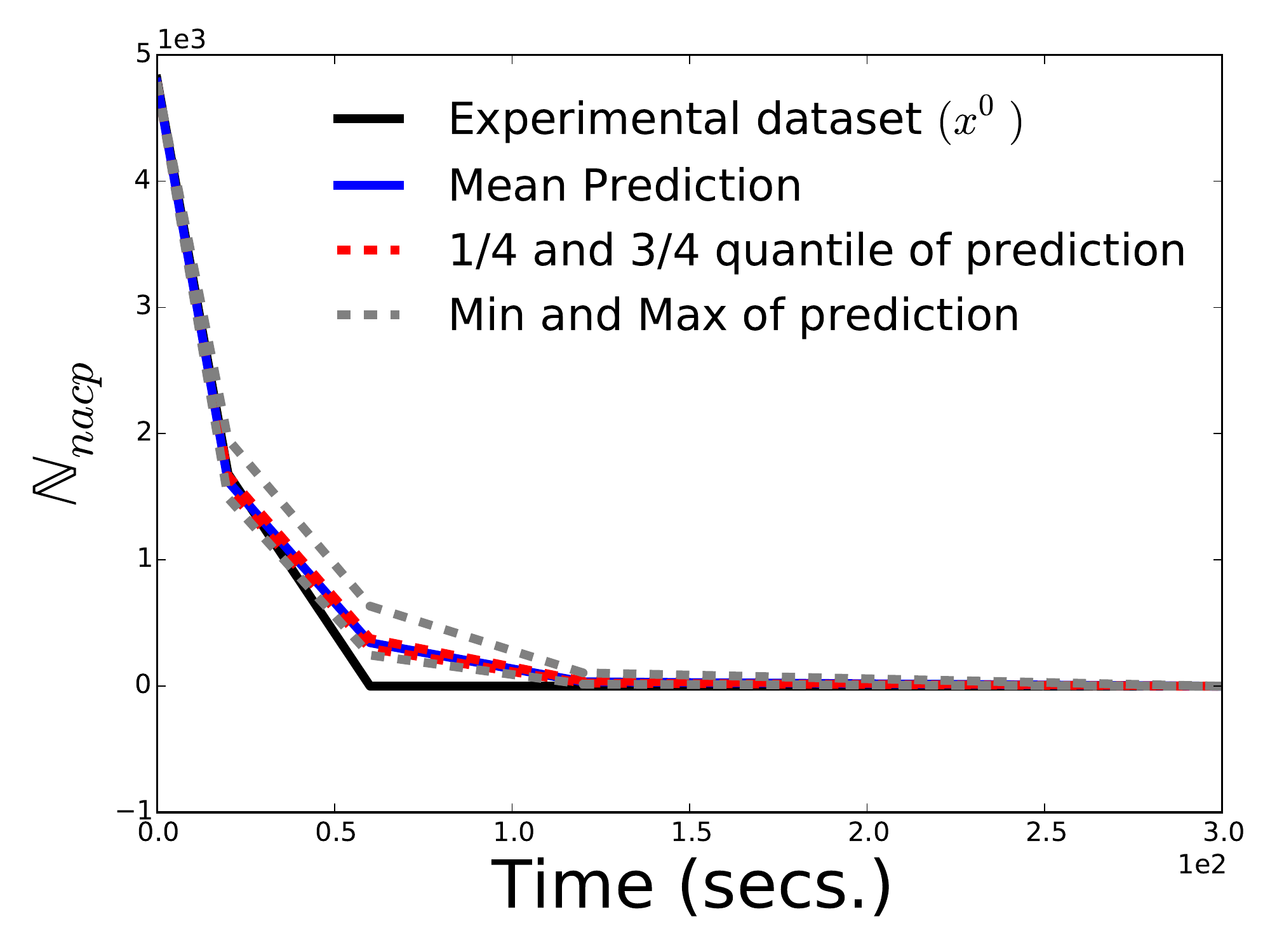}
\caption{Posterior Prediction Check: To validate the numerical model of the platelet deposition and the inference scheme we perform a posterior prediction check by simulating 100 datasets, each using a different parameter sample drawn from the posterior distribution. Here, we plot the experimental dataset (black-solid) used for inference, mean predicted dataset (blue-solid), $1/4$-th and $3/4$-th quantile (red-dashed), minimum and maximum (gray-dashed) of the predicted datasets at each timepoints.}
\label{fig:prediction_check_experimental}
\end{figure}

Next we do a Posterior predictive check to validate our model and inference scheme. The main goal here is to analyze the degree to which the experimental data deviate from the data generated from the inferred posterior distribution of the parameters. Hence we want to generate data from the model using parameters drawn from the posterior distribution. To do so, we first draw 100 parameter samples from the inferred approximate posterior distribution and simulate 100 data sets, each using a different parameter sample. We call this simulated dataset as the predicted dataset from our inferred posterior distribution and present the mean predicted dataset (blue-solid) compared with experimental dataset (black-solid) in Figure~\ref{fig:prediction_check_experimental}. Note that since we are dealing with the posterior distribution, we can also quantify uncertainty in our predictions. We plot the 1/4-th quantile, 3/4-th quantile (red-dashed), minimum and maximum (gray-dashed) of the predicted dataset at each timepoints to get a sense of uncertainty in the prediction. Here we see a very good agreement between the mean predicted dataset and the experimentally observed one, while the $1/4$-th and $3/4$-th quantile of the prediction being very tight. This shows a very good prediction performance of the numerical model of platelet deposition and the proposed inference scheme.   

\begin{figure}[htbp]
  \centering
\includegraphics[width=\textwidth]{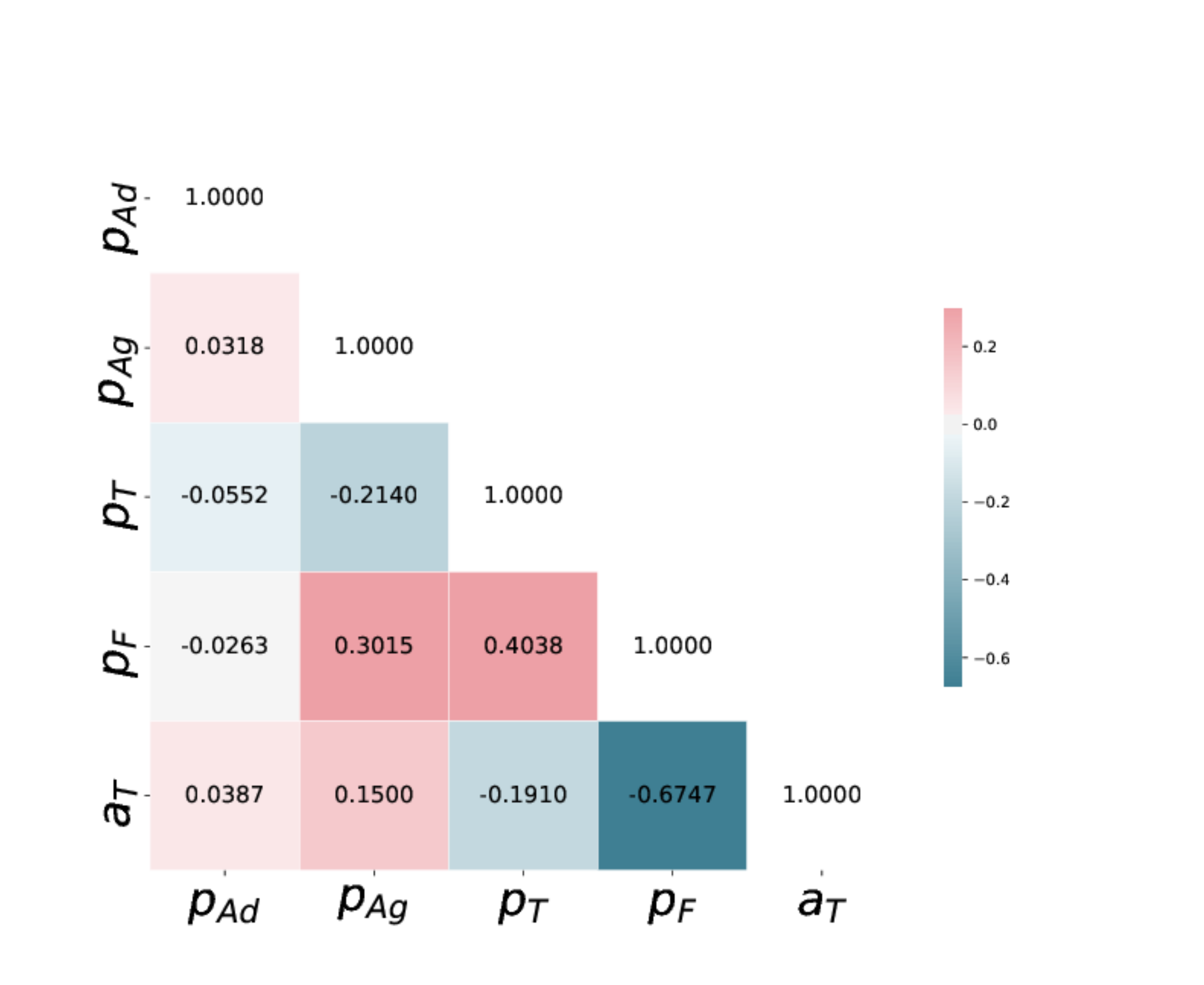}
  \caption{Posterior correlation matrix of $\left(\Pad, \Pg, \Pt, \Pf, \Ra  \right)$ computed from the 5000 i.i.d. samples drawn from the posterior distribution using SABC.}
  \label{fig:corr}
\end{figure}

Additionally, to point the strength of having a posterior distribution
for the parameters we compute and show the posterior correlation
matrix between the 5 parameters in Figure~\ref{fig:corr}, highlighting
a strong negative correlation between $(\Pf,\Ra)$, strong positive
correlations between $(\Pf,\Pg)$ and $(\Pf,\Pt)$. A detailed
investigation of these correlation structure would be needed to
understand them better, but generally they may point towards: a) the
stochastic nature of the considered model for platelet deposition and
b) the fact that the deposition process is an antagonistic or
synergetic combination of the mechanisms proposed in the model.

Note finally that the posterior distribution being the joint probability distribution of the 5 parameters, we can also compute any higher-order moments, skewness etc. of the parameters for a detailed statistical investigation of the natural phenomenon.     

\section{Conclusions}
\label{sec:conclusion}

Here, we have demonstrated that approximate Bayesian
computation (ABC) can be used to automatically explore the parameter
space of the numerical model simulating the deposition of platelets
subject to a shear flow as proposed in \cite{Chopard_2015,Chopard_2017}. We also notice the good agreement between the manually tuned parameters and the Bayes estimates, while saving us from  subjectivity and a tedious manual tuning. This approach can be applied patient per patient, in a systematic way, without the bias of a human
operator. In addition, the approach is computationally fast enough to
provide results in an acceptable time for contributing to a new
medical diagnosis, by giving clinical information that no other known method can
provide. The clinical relevance of this approach is still to be explored and our
next step will be to apply our approach at a personalized level, with
a cohort of patients with known pathologies. The possibility of designing new platelet functionality test as proposed here is the result of combining different techniques: advanced microscopic observation techniques, bottom-up numerical modeling and simulations, recent data-science development and high
performance computing (HPC).

Additionally, the ABC inference scheme provides us with a posterior distribution of the parameters given observed dataset, which is much more informative about the underlying process. The posterior correlations structure shown in Fig.~\ref{fig:corr} may not have a direct biophysical interpretation, though it illustrates some sort of underlying and unexplored stochastic mechanism for further investigation. Finally we note that, although the manual estimates achieve a very high posterior probability, they are different from the Bayes estimates learned using ABC. The departure reflects a different estimation of the quality of the match between experimental observation and simulation results. 
As the ABC algorithms are dependent on the choice of the summary statistics and the discrepancy measures, the parameter uncertainty quantified by SABC in Section~\ref{sec:results} or the Bayes estimates computed are dependent on the assumptions in Section~\ref{sec:postinf} regarding their choice. Fortunately there are recent works on automatic choice of summary statistics and discrepancy measures in ABC setup \citep{gutmann2017likelihood}, and incorporating some of these approaches in our inference scheme is a promising direction for future research in this area.

\section*{Author Contribution} 
Design of the research: RD, BC, AM; Performed research: RD; Experimental data collection: KZB, FD; writing of  the paper: RD, BC; Contribution to the  writing: AM, KZB, JL, FD, AM; design and coding of the numerical forward simulation model: BC, JL.

\section*{Ethics}
This study conforms with the Declaration of Helsinki and its protocol
was approved by the Ethics Committee of `CHU de Charleroi' (comité
d'éthique OM008). All volunteers gave their written informed consent.

\section*{Funding}
Ritabrata Dutta and Antonietta Mira are supported by Swiss National
Science Foundation Grant No. 105218\_163196 (Statistical Inference on
Large-Scale Mechanistic Network Models). We thanks CADMOS for
providing computing resources at the Swiss Super Computing Center. We
acknowledge partial funding from the European Union Horizon 2020
research and innovation programme for the CompBioMed project
(http://www.compbiomed.eu/) under grant agreement 675451.

\section*{Acknowledgments}
We thank Dr. Marcel Schoengens, CSCS, ETH Zurich for helps regarding.
We thank CHU Charleroi for supporting the experimental work used in
this study.

\bibliographystyle{plainnat}

\end{document}